\documentclass[aps,prd,onecolumn,showkeys,groupedaddress,showpacs,nofootinbib]{revtex4}
\usepackage{epsfig} \usepackage{amsmath} \usepackage{amsfonts}
\usepackage{amssymb} \usepackage{graphicx} \usepackage{colordvi}
\usepackage{psfrag}
 \usepackage{times}
 \usepackage{amsmath, amsthm, amssymb}
 \usepackage{makeidx}

  \makeindex
\makeatletter
\def\widebar{\accentset{{\cc@style\underline{\mskip10mu}}}}
\makeatother

\begin{document}
%\numberwithin{equation}{section}
\def\bib#1{[{\ref{#1}}]}
\title{\bf $f(R)$ cosmology  with torsion}

\author{S. Capozziello$^{1,2}$, R. Cianci$^{3}$, C. Stornaiolo$^{2}$, S.
Vignolo$^{3}$ }

\affiliation{$~^{1}$ Dipartimento di Scienze Fisiche,
Universit\`{a} ``Federico II'' di Napoli and $^{2}$INFN Sez. di
Napoli, Compl. Univ. Monte S. Angelo Ed. N, via Cinthia, I- 80126
Napoli (Italy)}

\affiliation{$^{3}$DIPTEM Sez. Metodi e Modelli Matematici,
Universit\`a di Genova,  Piazzale Kennedy, Pad. D - 16129 Genova
(Italy)}

\date{\today}

\begin{abstract}
$f(R)$-gravity with geometric torsion (not related to any spin
fluid) is considered in a cosmological context. We derive the
field equations in vacuum and in presence of perfect-fluid matter
and discuss the related cosmological models. Torsion vanishes in
vacuum for almost all arbitrary functions $f(R)$  leading to
standard General Relativity. Only for $f(R)=R^{2}$, torsion gives
contribution in the vacuum leading to an accelerated behavior .
When material sources are considered, we find that the torsion
tensor is different from zero even with spinless material sources.
This tensor is related to the logarithmic derivative of $f'(R)$,
which can be expressed also as a nonlinear function of the trace
of the matter energy-momentum tensor $\Sigma_{\mu\nu}$. We show
that the resulting equations for the metric can always be arranged
to yield effective Einstein equations.   When the homogeneous and
isotropic  cosmological models are considered,  terms originated
by torsion can lead  to accelerated expansion. This means that, in
$f(R)$ gravity, torsion can be a geometric source for
acceleration.
\end{abstract}


\maketitle

\section{Introduction}
$\Lambda$CDM  has recently assumed the role of a new Cosmological
Standard Model giving a coherent picture of the today observed
universe \cite{LambdaTest}. Although being the best fit to a wide
range of data, it suffers of several  theoretical shortcomings
\cite{LambdaRev} so it fails in tracking cosmic dynamics at every
redshift and fails in according observational cosmology to some
fundamental theory of physical interactions.  Among the defects of
this model, there is the lack of final probes, at fundamental
level, for dark energy and dark matter candidates (which should be
the $95\%$ of the energy-matter content of the universe!) which
frustrates the possibility to reduce $\Lambda$CDM  to some
self-consistent scheme, despite of the fact that it is a fair
"snapshot" of the present status of the universe.  These facts
motivate the search for other models, among which alternative
theories of gravity that should reproduce the successes of
$\Lambda$CDM but should be more appropriate in describing the
cosmological dynamics \cite{PR03,copeland}.

In particular, the large part of dark energy models  relies on the
implicit assumption that Einstein's General Relativity (GR) is the
correct theory of gravity indeed. Nevertheless, its validity on
large astrophysical and cosmological scales has never been tested
but only assumed \cite{will}, and it is therefore conceivable that
both cosmic speed up and missing matter, respectively the dark
energy and the dark matter,  are nothing else but signals of a
breakdown of GR at large scales. In other words, GR could fail in
giving self-consistent pictures both at ultraviolet scales (early
universe) and at infrared scales (late universe) also it is fairly
working at Solar System scales and in the weak field regime.

Staring from these considerations, a different possibility  could
be to better investigate  the gravitational sector and consider,
as source of the field equations, only the  observed (and probed
at fundamental level)  baryonic matter and radiation (photons and
neutrinos). A choice could be to take into account generic
functions $f(R)$ of the Ricci scalar $R$. The goal  should be to
match observational  data without considering exotic {\it dark}
ingredients, unless these are  found by means of  experiments at
fundamental level \cite{kleinert,noi}. This is the underlying
philosophy of the so-called \textit{$f(R)$-gravity} (see e.g.
\cite{copeland,odirev,GRGrev}).

These theories are receiving much attention in cosmology, since
they are  able to give rise to the accelerating expansion
\cite{noi} and it is possible to demonstrate that they play a
major role also at astrophysical scales (for  recent comprehensive
reviews see \cite{GRGrev,faraoni}). In fact, modifying the gravity
Lagrangian affects the gravitational potential in the low energy
limit. Provided that the modified potential reduces to the
Newtonian one on the Solar System scale, this implication could
represent an intriguing opportunity rather than a shortcoming for
$f(R)$ theories. In fact, a corrected gravitational potential
could offer the possibility to fit galaxy rotation curves without
the need of huge amounts of dark matter
\cite{noipla,mond,jcap,mnras,sobouti,salucci,mendoza}. In
addition, it is possible to work out a formal analogy between the
corrections to the Newtonian potential and the usually adopted
galaxy halo models which allow to reproduce dynamics and
observations without dark matter \cite{jcap}.

However, extending the gravitational Lagrangian could give rise to
several problems. These theories could have instabilities
\cite{instabilities-f(R)}, ghost\,-\,like behaviors
\cite{ghost-f(R)}, and they have to be matched with the low energy
limit experiments which quite fairly test GR.

In summary, it seems that the paradigm to adopt $f(R)$-gravity
leads to interesting results at cosmological, galactic and Solar
System scales but, up to now, no definite physical criterion has
been found to select the final $f(R)$ theory (or a class of
theories) capable of matching the data at all scales. Interesting
results have been achieved in this line of thinking
\cite{mimicking,Hu,Star,Odintsov1,tsuji} but the approaches are
all phenomenological and are not based on some fundamental
principle as the conservation or the invariance of some quantity
or some intrinsic symmetry of the theory. In any case, some
results are emerging in this direction and the presence of Noether
symmetries into dynamics seems to play an important role in
selecting physically interesting $f(R)$ theories
\cite{prado,antonio}.

Besides selection criteria, the "full" geometric sector of $f(R)$
gravity has to be investigated considering also the role of
torsion. Such an "ingredient"  has been firstly considered  by
Cartan and then by Sciama and Kibble in order to deal with spin in
GR (see \cite{hehl} for a review). Being spin as fundamental as
the mass of the particles, torsion was introduced in order to
complete the scheme that mass(energy) are the source of curvature
and spin is the source of torsion. Unfortunately, torsion in the
context of GR, seems not to produce models with observable effects
since the gravitational coupling is extremely weak in all the
torsion phenomena and only in the very early universe its effect
could have been significant.

However, it has been proven  that spin is not the only source for
torsion. As a matter of facts, torsion can be decomposed in three
irreducible tensors, with different properties. In
\cite{Capozziello:2001mq}, a systematic classification of these
different types of torsion and their possible sources was
discussed.

In two recent papers, the role of torsion in $f(R)$ gravity has
been considered in the framework of metric-affine formalism
\cite{cianci1,cianci2}. The field equations have been discussed in
empty space and in presence of perfect fluid matter taking into
account the analogy with the Palatini formalism \cite{magnano}. As a result, the
extra curvature and torsion degrees of freedom can be dealt as an
effective scalar field of fully geometric origin \cite{cianci1}.

From  a more formal viewpoint, $f(R)$ gravity with torsion can be
studied  in the framework of the ${\cal J}$-bundles formalism
\cite{cianci2}. Such an approach is particularly useful since the
components of the torsion and curvature tensors can be chosen as
fiber ${\cal J}$-coordinates on the bundles and then the
symmetries and the conservation laws of the theory can be easily
achieved. Also in this case, field equations of $f(R)$-gravity
have been studied in empty space and in presence of various forms
of matter as Dirac fields, Yang--Mills fields and spin perfect
fluid. Such fields enlarge the jet bundles framework and
characterize the dynamics.

In this paper,  we discuss the cosmological applications of
$f(R)$-gravity with torsion,  considering the possibility that the
whole dark side of the universe (dark matter and dark energy)
could be geometrically interpreted by curvature and torsion.
Examples in which repulsive gravity and clustered structures could
be implemented considering torsion are present in literature
\cite{hehl,gasperini,szydlowski} but, in that cases, GR has been
adopted and the whole dark sector has not been addressed.

The layout of the paper is the following. In Sec.II, we give the
definitions and conventions for torsion tensor and construct the
in the main geometrical quantities in $U_4$ spacetime. Field
equations in vacuum are derived in Sec.III. In Sec.IV, we derive
the cosmological equations in vacuum and find out some interesting solutions
where cosmological term is given by the trace of torsion tensor.
Field equations, in presence of standard fluid matter, are
discussed in Sec.V, while, in Sec.VI, also spin is considered. A
relevant discussion is devoted to the fact that torsion
contribution can be dealt under the standard of a scalar field. In
Sec.VII, we consider weak and strong energy conditions in $f(R)$
gravity with torsion. Cosmological models in $f(R)$ gravity with
torsion and  exact solutions are discussed in Sec.VIII. Discussion
and conclusions are drawn in Sec.IX.

\section{Torsion tensor and invariant quantities in $U_4$}

A 4D-differential manifold equipped with torsion is defined in a
$U_4$ space, while standard torsionless Riemannian manifolds are
defined in $V_4$. In literature, there are several  definitions
for torsion and  quantities related to it. For a summary, see
\cite{Capozziello:2001mq}. The conventions adopted in this paper
are those in \cite{hehl}.

The torsion tensor $S_{ab}^{\phantom{ab}c}$ is defined as the
antisymmetric part of connection in a coordinate basis
\begin{equation}\label{3a}
    S_{\mu\nu}^{\phantom{\mu\nu}\lambda}=
    \Gamma_{[{\mu\nu}]}^{\phantom{[\mu\nu]}\lambda}\equiv \frac{1}{2}
    \left(\Gamma_{\mu\nu}^{\phantom{\mu\nu}\lambda}-\Gamma_{\nu\mu}^{\phantom{\nu\mu}\lambda}\right)
\end{equation}
a connection with torsion and metric-compatible
($\nabla_{\lambda}g_{\mu\nu}=0$) has the form
\begin{equation}\label{4a}
  \Gamma_{\mu\nu}^{\phantom{\mu\nu}\lambda}= \left\{_{\mu\nu}^{\phantom{\mu\nu}\lambda}\right\}  + S_{\mu\nu}^{\phantom{\mu\nu}\lambda} -  S_{\nu\phantom{\lambda}\mu}^{\phantom{\nu}\lambda} +  S_{\phantom{\lambda}\mu\nu}^{\lambda}=\left\{_{\mu\nu}^{\phantom{\mu\nu}\lambda}\right\}  - K_{\mu\nu}^{\phantom{\mu\nu}\lambda}
\end{equation}
where $\left\{_{\mu\nu}^{\phantom{\beta}\lambda}\right\}$ are the
Christoffel symbols and
\begin{equation}\label{5a}
    K_{\mu\nu}^{\phantom{\mu\nu}\lambda}=-S_{\mu\nu}^{\phantom{\mu\nu}\lambda} + S_{\nu\phantom{\lambda}\mu}^{\phantom{\nu}\lambda} - S_{\phantom{\lambda}\mu\nu}^{\lambda}
\end{equation}
is  the contorsion tensor. Another  combination of torsion tensor,
often used in the calculations, is the modified torsion defined by
the following relation
\begin{equation}\label{6a}
     T_{\mu\nu}^{\phantom{\mu\nu}\lambda} = S_{\mu\nu}^{\phantom{\mu\nu}\lambda}+\delta_{\mu}^{\lambda}S_{\nu}- \delta_{\nu}^{\lambda}S_{\mu}
\end{equation}
where

\begin{equation}\label{7a}
 S_{\mu}=   S_{\mu\nu}^{\phantom{\mu\nu}\nu}
\end{equation}
is  the torsion trace-vector.

Here, we shall consider a vectorial torsion
\cite{Capozziello:2001mq} of the form
 \begin{equation}\label{a1}
 S_{\mu\nu}^{\phantom{\mu\nu}\lambda}=A_{[\mu}\delta_{\nu]}^{\lambda}\,.
 \end{equation}
Its trace is
\begin{equation}\label{a2}
   S_{\mu}= S_{\mu\nu}^{\phantom{\mu\nu}\nu}=\frac{3}{2}A_{\mu}
\end{equation}
so the vectorial torsion takes the form
 \begin{equation}\label{a3}
 S_{\mu\nu}^{\phantom{\mu\nu}\lambda}=\frac{2}{3}S_{[\mu}\delta_{\nu]}^{\lambda}\,
.
 \end{equation}
In the following, we will find convenient to express  the field
equations   in terms of  one of these quantities. To this purpose,
let us give here the relations  among  torsion, modified torsion,
contorsion, and their respective traces. From Eq.(\ref{6a}), it is

\begin{equation}
 T_{\mu\nu}^{\phantom{\mu\nu}\lambda}  =
-\frac{2}{3}(\delta_{\mu}^{\lambda}S_{\nu}-
\delta_{\nu}^{\lambda}S_{\mu})=-
2S_{\mu\nu}^{\phantom{\mu\nu}\lambda}\,,
\end{equation}
and then
\begin{equation}\label{a7}
     T_{\mu}=-2S_{\mu}\ \ \ \ \ \ \ \textrm{or}\ \ \ \ \ \ S_{\mu}=- \frac{1}{2}T_{\mu}
\end{equation}
where the modified torsion can be expressed in the following form
 \begin{equation}\label{a8}
T_{\mu\nu}^{\phantom{\mu\nu}\lambda}=\frac{2}{3}T_{[\mu}\delta_{\nu]}^{\lambda}\,
.
 \end{equation}
The Riemann tensor can  be decomposed in a part depending only on
the Christoffel symbols and a part which contains the contorsion
tensor. Using Eq.(\ref{a8}) in Eq.(\ref{5a}),  we obtain
\begin{equation}\label{a10}
K_{\mu\nu}^{\phantom{\mu\nu}\lambda}=\frac{1}{2}\left(
T_{\mu\nu}^{\phantom{\mu\nu}\lambda}-T_{\nu\phantom{\lambda}\mu}^{\phantom{\nu}\lambda}
+ T_{\phantom{\lambda}\mu\nu}^{\lambda}\right)\, .
\end{equation}
Finally,  substituting (\ref{a8}) in (\ref{a10}), we get
\begin{equation}\label{a11}
K_{\mu\nu}^{\phantom{\mu\nu}\lambda}=\frac{1}{3}\left(
T^{\lambda}g_{\mu\nu}-\delta_{\mu}^{\lambda}T_{\nu}\right)
\end{equation}
and
\begin{equation}\label{a12}
     K^{\lambda}=g^{\mu\nu}K_{\mu\nu}^{\phantom{\mu\nu}\lambda}=T^{\lambda}.
\end{equation}

Furthermore, starting from the standard definition of the Riemann
tensor
\begin{equation}\label{b1}
     R^{\alpha}_{\phantom{\alpha}\beta\mu\nu}=\partial_{\mu}\Gamma^{\alpha}_{\nu\beta}-\partial_{\nu}\Gamma^{\alpha}_{\mu\beta}+
     \Gamma^{\sigma}_{\nu\beta}\Gamma^{\alpha}_{\mu\sigma}-
     \Gamma^{\sigma}_{\mu\beta}\Gamma^{\alpha}_{\nu\sigma}\,,
\end{equation}
one can insert the decomposition of the connection in the
Christoffel and contorsion parts
\begin{equation}\label{b2}
\Gamma^{\alpha}_{\nu\beta}=\left\{^{\,\alpha}_{\nu\beta}\right\}-K^{\phantom{\nu\beta}\alpha}_{\nu\beta}\,.
\end{equation}
In this way,  the Riemann tensor results decomposed in a first
term obtained by the Christoffel connection and its derivative and
a second term given by the contorsion and its derivative.
$$   R^{\alpha}_{\phantom{\alpha}\beta\mu\nu}=\partial_{\mu}\left(\left\{^{\,\alpha}_{\nu\beta}\right\}-K^{\phantom{\nu\beta}\alpha}_{\nu\beta}
     \right)-\partial_{\nu}\left(\left\{^{\,\alpha}_{\mu\beta}\right\}-K^{\phantom{\mu\beta}\alpha}_{\mu\beta}\right)$$
\begin{equation}\label{b3}
  +\left(\left\{^{\,\sigma}_{\nu\beta}\right\}-K^{\phantom{\nu\beta}\sigma}_{\nu\beta}\right)\left(\left\{^{\,\alpha}_{\mu\sigma}\right\}-K^{\phantom{\mu\sigma}\alpha}_{\mu\sigma}\right)-
    \left(\left\{^{\,\sigma}_{\mu\beta}\right\}-K^{\phantom{\mu\beta}\sigma}_{\mu\beta}\right)
    \left(\left\{^{\,\alpha}_{\nu\sigma}\right\}-K^{\phantom{\nu\sigma}\alpha}_{\nu\sigma}\right)\,.
\end{equation}
Using the covariant derivative without contorsion
$\widetilde{\nabla}  $, the previous expression takes the form
\begin{equation}\label{b4}
   R^{\alpha}_{\phantom{\alpha}\beta\mu\nu}= R^{\alpha}_{\phantom{\alpha}\beta\mu\nu}(\{\})+\widetilde{\nabla}_{\nu} K^{\phantom{\mu\beta}\alpha}_{\mu\beta}-  \widetilde{\nabla}_{\mu}K^{\phantom{\nu\beta}\alpha}_{\nu\beta}
    +K^{\phantom{\nu\beta}\sigma}_{\nu\beta}K^{\phantom{\mu\sigma}\alpha}_{\mu\sigma}
  -K^{\phantom{\mu\beta}\sigma}_{\mu\beta}K^{\phantom{\nu\sigma}\alpha}_{\nu\sigma}\,
  .
\end{equation}
The corresponding Ricci tensor  and the curvature scalar are
respectively

\begin{equation}\label{b5}
    R_{\beta\mu}\equiv R^{\alpha}_{\phantom{\alpha}\beta\alpha\mu}=
    R_{\beta\mu}(\{\})
    + \widetilde{\nabla}_{\mu}K^{\phantom{\alpha\beta}\alpha}_{\alpha\beta}
    - \widetilde{\nabla}_{\alpha}K^{\phantom{\mu\beta}\alpha}_{\mu\beta}
    +K^{\phantom{\mu\beta}\sigma}_{\mu\beta}K^{\phantom{\alpha\sigma}\alpha}_{\alpha\sigma}
    -K^{\phantom{\alpha\beta}\sigma}_{\alpha\beta}K^{\phantom{\mu\sigma}\alpha}_{\mu\sigma}
    \end{equation}

and

\begin{equation}\label{b6}
    R\equiv g^{\mu\beta}R_{\beta\mu}=
    R_{\beta\mu}(\{\})
    +g^{\mu\beta} \widetilde{\nabla}_{\mu}K^{\phantom{\alpha\beta}\alpha}_{\alpha\beta}
    - g^{\mu\beta}\widetilde{\nabla}_{\alpha}K^{\phantom{\mu\beta}\alpha}_{\mu\beta}
    +g^{\mu\beta}K^{\phantom{\mu\beta}\sigma}_{\mu\beta}K^{\phantom{\alpha\sigma}\alpha}_{\alpha\sigma}
    -g^{\mu\beta}K^{\phantom{\alpha\beta}\sigma}_{\alpha\beta}K^{\phantom{\mu\sigma}\alpha}_{\mu\sigma}\,.
\end{equation}
If torsion is  in the vectorial form, the Ricci tensor and the
scalar curvature take the forms
\begin{equation}\label{b7}
  R_{\beta\mu}= R_{\beta\mu}(\{\})- \frac{2}{3}
  \widetilde{\nabla}_{\mu}T_{\beta}-\frac{1}{3}g_{\mu\beta}
  \widetilde{\nabla}_{\lambda}T^{\lambda}-\frac{2}{9}g_{\mu\beta} T^{\lambda}T_{\lambda}+\frac{2}{9} T_{\beta}T_{\mu}
\end{equation}
\begin{equation}\label{b8}
   R=R(\{\}) - 2\widetilde{\nabla}_{\lambda}T^{\lambda} -
    \frac{2}{3}T^{\lambda}T_{\lambda}\,,
\end{equation}
which will be widely used in the following discussion.

\section{Field equations in the vacuum}
Let us start our considerations deriving the field equations in
vacuum. The action for the gravitational part is
\begin{equation}\label{1}
     \mathcal{A}=\int\sqrt{-g}f(R)d^{4}x\,.
\end{equation}
It is instructive to report the variational principle step by step
in order to put in evidence the differences with respect to a
variation in $V_4$. It is

\begin{equation}\label{2}
    \delta\mathcal{A}=\int\left[ \delta\sqrt{-g} f(R) + \sqrt{-g}\delta f(R)
    \right]d^{4}x
\end{equation}
\begin{equation}\label{3}
    \phantom{\delta\mathcal{A}}=\int\sqrt{-g}\left[f'(R)R_{\mu\nu}-\frac{1}{2}g_{\mu\nu}f(R)\right]\delta
    g^{\mu\nu}d^{4}x +\int\sqrt{-g}f'(R)g^{\mu\nu}\delta R_{\mu\nu}d^{4}x
\end{equation}

    $$ \phantom{\delta\mathcal{A}}= \int\sqrt{-g}\left[f'(R)R_{\mu\nu}-\frac{1}{2}g_{\mu\nu}f(R)\right]\delta g^{\mu\nu}d^{4}x +\int\sqrt{-g}f'(R)g^{\mu\nu} \left[
2\nabla_{[\sigma}\delta\Gamma_{\mu]\nu}^{\sigma} \right.$$

\begin{equation}\label{4}
 \left.   + 2 S_{\sigma\mu}^{\phantom{\sigma\mu}\lambda}\delta \Gamma_{\lambda\nu}^{\phantom{\lambda\nu}\sigma}
 \right ]d^{4}x\,.
\end{equation}
In a space equipped with torsion,  the following property holds
\begin{equation}\label{5}
    \partial (P\delta Q)= P\nabla\delta Q +( \breve{\nabla}P) \delta
    Q\,,
\end{equation}
where $P\delta Q$ is a vectorial density and
$\breve{\nabla}_{\mu}=
\nabla_{\mu}+2S_{\mu\nu}^{\phantom{\mu\nu}\nu} $. If the variation
$\delta Q$  is zero on the boundary or at infinity, we find
\begin{equation}\label{6}
     P\nabla\delta Q =- (\breve{\nabla}P) \delta Q\,.
\end{equation}
The variation assumes the form
 $$ \delta\mathcal{A}= \int\sqrt{-g}\left[f'(R)R_{\mu\nu}-\frac{1}{2}g_{\mu\nu}f(R)\right]\delta g^{\mu\nu}d^{4}x $$

\begin{equation}\label{7}
+\int\sqrt{-g}\left[\breve{\nabla}_{\lambda} (f'(R)g^{\mu\nu})
 \left(\delta^{\lambda}_{\mu}\delta^{\rho}_{\sigma}-\delta^{\lambda}_{\sigma}\delta^{\rho}_{\mu}\right)
\delta\Gamma_{\rho\nu}^{\sigma} +
2f'(R)g^{\mu\nu}S_{\sigma\mu}^{\phantom{\sigma\mu}\rho}\delta\Gamma_{\rho\nu}^{\sigma}\right]\,.
d^{4}x
\end{equation}
Giving the explicit expression for  $\breve{\nabla}$, we obtain
 $$ \delta\mathcal{A}= \int\sqrt{-g}\left[f'(R)R_{\mu\nu}-\frac{1}{2}g_{\mu\nu}f(R)\right]\delta g^{\mu\nu}d^{4}x $$
\begin{equation}\label{8}
+\int\sqrt{-g}\left[ g^{\mu\nu}\partial_{\lambda} (f'(R))
 \left(\delta^{\lambda}_{\mu}\delta^{\rho}_{\sigma}-\delta^{\lambda}_{\sigma}\delta^{\rho}_{\mu}\right)+
2f'(R)g^{\mu\nu}T_{\sigma\mu}^{\phantom{\sigma\mu}\rho}\right]\delta\Gamma_{\rho\nu}^{\sigma}
d^{4}x
\end{equation}
where  $T_{\sigma\mu}^{\phantom{\sigma\mu}\rho}$ is the above
modified torsion tensor. Finally, the variation of the connection,
expressed in terms of the metric and the contorsion, is
\begin{equation}\label{9}
\delta\Gamma_{\rho\nu}^{\sigma}=\frac{1}{2}g^{\sigma\lambda}\left(\nabla_{\nu}\delta
g_{\lambda\rho}+\nabla_{\rho}\delta
g_{\nu\lambda}-\nabla_{\lambda}\delta g_{\rho\nu}\right)+
g_{\gamma\delta} \delta
K_{\alpha\beta}^{\phantom{\alpha\beta}\delta}\,.
\end{equation}
By substituting Eq.(\ref{9}) in (\ref{8}), we obtain
\begin{equation}\label{10}
   U_{\sigma\mu}^{\phantom{\sigma\mu}\rho}\equiv  \frac{\delta \sqrt{-g}f(R)}{\delta
    K_{\alpha\beta}^{\phantom{\alpha\beta}\delta}}=g^{\mu\nu}\partial_{\lambda} (f'(R))
 \left(\delta^{\lambda}_{\mu}\delta^{\rho}_{\sigma}-\delta^{\lambda}_{\sigma}\delta^{\rho}_{\mu}\right)+
2f'(R)g^{\mu\nu}T_{\sigma\mu}^{\phantom{\sigma\mu}\rho}
=0
\end{equation}
In general, $f(R)$ theories, in metric formalism, present fourth
order terms in the field equations. In  this case, such terms are
absorbed in the torsion components and then

\begin{equation}\label{11}
    \frac{\delta \sqrt{-g}f(R)}{\delta
    g_{\mu\nu}}=- f'(R) R^{\mu\nu} + \frac{1}{2} g^{\mu\nu}f(R)+
    \breve{\nabla}_{\lambda}
    (U^{\mu\nu\lambda}-U^{\nu\lambda\mu}+U^{\lambda\mu\nu})=0\,.
\end{equation}
The divergence in Eq.(\ref{11}) is  zero  because of Eq.(\ref{10})
and then the field equations in vacuum are
\begin{equation}\label{13}
     f'(R) R^{\mu\nu} - \frac{1}{2} g^{\mu\nu}f(R)=0
\end{equation}
and
\begin{equation}\label{14}
f'(R) T_{\sigma\mu}^{\phantom{\sigma\mu}\rho}+\frac{1}{2}\,
\partial_{\lambda} (f'(R))
 \left(\delta^{\lambda}_{\mu}\delta^{\rho}_{\sigma}-\delta^{\lambda}_{\sigma}\delta^{\rho}_{\mu}\right)=0\,
 .
\end{equation}
Taking into account the trace of Eq.(\ref{13})
\begin{equation}\label{15}
     f'(R) R- 2 f(R)=0\, ,
\end{equation}
and substituting it in  Eq. (\ref{13}), it follows
\begin{equation}\label{16}
    f'(R)\left( R_{\mu\nu}-\frac{1}{4}g_{\mu\nu}R\right)=0
\end{equation}
The set of Eqs.(\ref{15}) and (\ref{16}) is equivalent to
Eq.(\ref{13}).

In general, Eq.(\ref{15})  is an algebraic or transcendent
equation for $R$ which is satisfied if  and only if $R$ is a
constant. This means, from  Eq.(\ref{14}),  that torsion is
identically zero and then we have. The resulting spacetime may be non-trivial, say a  Schwarzschild-de Sitter. The trivial case may be de Sitter or anti-de Sitter, according to the sign of $\Lambda$, i.e.
%|
\begin{equation}\label{16b}
     R_{\mu\nu}= \Lambda g_{\mu\nu}\,.
\end{equation}

Only for $f(R)=\alpha R^{2}$,  Eq.(\ref{15}) is an identity and
$R$ is not necessarily a constant. In this case, torsion is
different from zero in the vacuum. By substituting  $f(R)=\alpha
R^{2}$ in Eq.(\ref{14}), we have
\begin{equation}\label{17}
 T_{\sigma\mu}^{\phantom{\sigma\mu}\rho}+
\frac{\partial_{\lambda}  R}{R}
\frac{1}{2} \left(\delta^{\lambda}_{\mu}\delta^{\rho}_{\sigma}-\delta^{\lambda}_{\sigma}\delta^{\rho}_{\mu}\right)=0\, .
\end{equation}
which  can be reduced  to
\begin{equation}\label{3.19}
   T_{\sigma}=T_{\sigma\mu}^{\phantom{\sigma\mu}\mu}=\frac{3}{2}\frac{\partial_{\sigma}R}{R}\,.
\end{equation}
This equation, together with
\begin{equation}\label{18}
    R_{\mu\nu}-\frac{1}{4}g_{\mu\nu}R=0
\end{equation}
is a system of equations for the gravitational field.

We can decompose the curvature terms in Eqs. (\ref{18}) and
(\ref{3.19}) in a part given by the Christoffel symbols and a part
 depending on torsion. The resulting equation  is
\begin{equation}\label{3.20}
     R_{\mu\nu}(\{\})- \frac{1}{4}g_{\mu\nu}R(\{\})-
     \frac{2}{3}\widetilde{\nabla}_{\nu}T_{\mu}+\frac{1}{6}g_{\mu\nu}\widetilde{\nabla}^{\sigma}T_{\sigma}+
     \frac{2}{9}T_{\mu}T_{\nu}-\frac{1}{18}g_{\mu\nu}T^{\lambda}T_{\lambda}=0
\end{equation}
where the symbols $\widetilde{\nabla}$ and $R(\{\})$ represent,
respectively, the covariant derivative and the scalar curvature
given by the Christoffel symbols. Similarly, the Ricci scalar $R$
results
\begin{equation}\label{3.21}
   R=R(\{\}) - 2\widetilde{\nabla}_{\lambda}T^{\lambda}-
    \frac{2}{3}T^{\lambda}T_{\lambda}\,.
\end{equation}
On the other hand,    the vector component of torsion $T_{\sigma}$
satisfies the following  differential equation
\begin{equation}\label{3.23}
 \frac{3}{2}\,\partial_{\sigma}\left(R(\{\})-2\widetilde{\nabla}_{\lambda}T^{\lambda}
- \frac{2}{3}T_{\lambda}T^{\lambda}   \right)
=\left(R(\{\})-2\widetilde{\nabla}_{\lambda}T^{\lambda}
- \frac{2}{3}T_{\lambda}T^{\lambda}   \right)T_{\sigma}
\end{equation}
which is a  propagation equation for $T_{\sigma}$. Finally,
applying the contracted  Bianchi identities, adding and
subtracting $R(\{\})/4$ in   Eq.(\ref{3.20})  and taking the
covariant divergence with respect to the $\widetilde{\nabla}$
derivative, it follows that
\begin{equation}\label{3.24}
     \frac{2}{3}g^{\nu\lambda}\widetilde{\nabla}_{\nu}\widetilde{\nabla}_{ \lambda}T_{\mu} - \frac{2}{9}\widetilde{\nabla}_{\nu}(T_{\mu}T^{\nu})-\frac{1}{4}\partial_{\mu}R-\frac{1}{6}\partial_{\mu}\widetilde{\nabla}_{\sigma}T^{\sigma}
+\frac{1}{18}\partial_{\mu}(T_{\lambda}T^{\lambda})=0\,,
\end{equation}
which we will be useful for the discussion below.

\section{Cosmology with torsion in  vacuum}

In order to develop our cosmological considerations, let us take
into account a Friedmann-Robertson-Walker metric of the type
\begin{equation}\label{3.25}
     ds^{2}=dt^{2}-\frac{a^{2}(t)}{\left(1+\frac{k}{4}r^{2}\right)^{2}}\left[dx^{2}+dy^{2}+dz^{2}\right]\,.
\end{equation}
The scalar (Christoffel) curvature in such a metric is
\begin{equation}\label{3.22}
   R(\{\})= -6\left[ \frac{\ddot{a}}{a}+\left( \frac{\dot{a}}{a}\right)^{2}+\frac{k\,}{a^{2}}\right] =
   -6\left(\dot{H}+2H^{2}+\frac{k\;}{a^{2}}\right)\,.
\end{equation}
Being the system dependent only on $t$,  from  Eq.(\ref{3.23}),
only the $T_{0}=T$ component of  $T_{\sigma}$ is different from
zero. Eqs.(\ref{3.20}) and  (\ref{3.22}), after some algebraic
manipulations, give the following cosmological equations
\begin{equation}\label{3.27}
     \frac{3}{2}\frac{d}{dt}\left(6\dot{H}+12H^{2}+6\frac{k}{a^{2}}+2\dot{T}+6HT+\frac{2}{3}T^{2}\right) =
     \left(6\dot{H}+12H^{2}+6\frac{k}{a^{2}}+2\dot{T}+6HT+\frac{2}{3}T^{2}\right)T\,,
\end{equation}
\begin{equation}\label{3.28}
    \frac{\ddot{a}}{a}-\frac{\dot{a}^{2}}{a^{2}}-\frac{k}{a^{2}}=-\frac{1}{3}\dot{T}+
    \frac{1}{3}HT+\frac{1}{9}T^{2}\,,
\end{equation}

which, in terms of the Hubble parameter $H$, can be reduced
respectively to
\begin{equation}\label{3.29}
    3\ddot{H}+12H\dot{H}-6\frac{k}{a^{2}}H+\ddot{T}+\dot{H}T+3H\dot{T}=4H^{2}T+2\frac{k}{a^{2}}T+2HT^{2}+\frac{2}{9}T^{3}
\end{equation}
and

\begin{equation}\label{3.30}
     \dot{H}-\frac{k}{a^{2}}=-\frac{1}{3}\dot{T}+\frac{1}{3}HT+\frac{1}{9}T^{2}
\end{equation}
The  Bianchi identities (\ref{3.24}) gives
\begin{equation}\label{3.31}
     3\ddot{T}-2T\dot{T}+12H\dot{T}-12H^{2}T -3H\dot{T}
-3\dot{H}T-4HT^{2}+9\ddot{H}+36H\dot{H}-9H\frac{k}{a^{2}}=0
\end{equation}
while the spatial components reduce to an algebraic identity. Let
us find a solution for the system (\ref{3.29}), (\ref{3.30})
(\ref{3.30}) in the physically interesting case $k=0$
corresponding to a spatially flat universe. Substituting $\dot{H}$
in (\ref{3.31}), it reduces to
 \begin{equation}\label{3.32}
     \frac{d}{dt}\left(H+\frac{1}{3}T\right)^{2}=\frac{2}{3}\left(H+\frac{1}{3}T\right)^{2}T\,,
 \end{equation}
and then we have two cases
\begin{eqnarray}
 \label{3.33} H+\frac{1}{3}T &=& 0 \,,\\
 \label{3.34} \frac{d}{dt}\left(H+\frac{1}{3}T\right) &=&
 \frac{1}{3}\left(H+\frac{1}{3}T\right)T\,,
\end{eqnarray}
In the first case, we have the solution
\begin{equation}
\label{exp}
a(t)=a_0\exp\left(-\frac{1}{3}Tt\right)
\end{equation}
In the second case, we obtain
\begin{equation}\label{3.35}
    H=H_{0}\exp\left(\frac{T}{3}t\right) -\frac{ T}{3}
\end{equation}
which gives a cosmological expansion driven by torsion:
\begin{equation}\label{3.36}
     a(t)=a_{0}\exp\left[-\frac{T}{3}t+A_{0}\exp\left(\frac{T}{3}t\right)\right]
\end{equation}
being $H_{0}$, $a_{0}$ and $A_{0}$ arbitrary integration
constants. This situation is particularly interesting in view of
considering vectorial torsion as the  geometrical source of
cosmological accelerated expansion also without assuming an
additional spin fluid \cite{szydlowski}. It is important to point out at this point
that solutions (\ref{exp}) and (\ref{3.36}) are derived in the particular case of $T=$constant but they can be derived also for $T=T(t)$. In this general case, $T$ has to be substituted with $\bar{T}=\int T(t)dt$.
These results are clear indications that geometric torsion can be an effective source for inflation and, in general, accelerated expansion.

\section{Field equations in presence of perfect-fluid matter}
A more realistic situation is considering, in the  action
(\ref{1}), the presence of a perfect fluid matter Lagrangian
density. We have
\begin{equation}\label{2.1}
     \mathcal{A}=\int \sqrt{-g}[f(R)+ \mathcal{L}_{m}]d^{4}x\,.
\end{equation}
By the sake of simplicity, let us consider first $\mathcal{L}_{m}$
not containing torsion terms. The corresponding matter fluid
energy-momentum tensor is $\Sigma^{\mu\nu}$ and
$\Sigma=g_{\mu\nu}\Sigma^{\mu\nu}$ its trace. The field equations
are then

\begin{equation}\label{2.2}
     f'(R) R_{\mu\nu} - \frac{1}{2} g_{\mu\nu}f(R)= \Sigma_{\mu\nu}
\end{equation}
and
\begin{equation}\label{2.3}
f'(R) T_{\sigma\mu}^{\phantom{\sigma\mu}\rho}+\frac{1}{2}\,
\partial_{\lambda} (f'(R))
 \left(\delta^{\lambda}_{\mu}\delta^{\rho}_{\sigma}-\delta^{\lambda}_{\sigma}\delta^{\rho}_{\mu}\right)=0\,
 .
\end{equation}
The trace of Eq.(\ref{2.2})
\begin{equation}\label{2.4}
     f'(R) R- 2 f(R)=\Sigma\, ,
\end{equation}
gives a non-linear relation between the curvature scalar $R$ and
$\Sigma$ which is
\begin{equation}\label{4.5}
R=F(\Sigma)\,,
\end{equation}
and then Eq.(\ref{2.2}) and (\ref{2.3}) can be rewritten as
\begin{equation}\label{4.6}
R_{\mu\nu}-\frac{1}{4}g_{\mu\nu}R=
\frac{1}{f'(F(\Sigma))}\left(\Sigma_{\mu\nu}-\frac{1}{4}g_{\mu\nu}\Sigma\right)
\end{equation}

\begin{equation}\label{4.7}
 T_{\sigma\mu}^{\phantom{\sigma\mu}\rho}=-\frac{1}{2}\,
\frac{\partial_{\lambda}  f'(F(\Sigma))}{f'(F(\Sigma))}
 \left(\delta^{\lambda}_{\mu}\delta^{\rho}_{\sigma}-\delta^{\lambda}_{\sigma}\delta^{\rho}_{\mu}\right)\,.
\end{equation}
It is worth noting that the equation for torsion is now an
``algebraic" expression. Torsion is present as a sort of
gravitational coupling being $f'(F(\Sigma))$. Eq.(\ref{4.6}) can
be written in an Einstein form by adding and subtracting  $-
\frac{1}{4} g_{\mu\nu}R$ and then using Eq.(\ref{4.5}). We obtain
\begin{equation}\label{4.8}
R_{\mu\nu}-\frac{1}{2}g_{\mu\nu}R=
\frac{1}{f'(F(\Sigma))}\left(\Sigma_{\mu\nu}-\frac{1}{4}g_{\mu\nu}\Sigma\right)
- \frac{1}{4}g_{\mu\nu} F(\Sigma)\,.
\end{equation}
By  using Eqs.(\ref{3.20}) and (\ref{3.21}), we can decompose the
Ricci tensor and the curvature scalar in their Christoffel and the
torsion dependent terms
\begin{eqnarray}
\nonumber  R_{\mu\nu}(\{\})-\frac{1}{2}g_{\mu\nu}R(\{\}) &=& \frac{1}{f'(F(\Sigma))}\left(\Sigma_{\mu\nu}-\frac{1}{4}g_{\mu\nu}\Sigma\right) - \frac{1}{4}g_{\mu\nu} F(\Sigma) +\frac{2}{3}\widetilde{\nabla}_{\mu}T_{\nu}   \\
  &&-\frac{2}{3} g_{\mu\nu} \widetilde{\nabla}_{\lambda}T^{\lambda} \  -\frac{2}{9}   T_{\mu}T_{\nu} -\frac{1}{9}g_{\mu\nu} T_{\lambda}T^{\lambda}. \label{4.9}
\end{eqnarray}
By taking the trace of Eq.(\ref{4.7})
\begin{equation}\label{4.10}
    T_{\sigma}=
    \frac{3}{2}\frac{\partial_{\sigma}\varphi}{\varphi}\,,
\end{equation}
where we define the auxiliary scalar field
\begin{equation}\label{4.11}
     \varphi = f'(F(\Sigma))\,,
\end{equation}
we obtain
\begin{eqnarray}
\nonumber G_{\mu\nu}(\{\}) &=& \frac{1}{\varphi} \Sigma_{\mu\nu}-\frac{3}{2}\frac{1}{\varphi^{2}}  \partial_{\mu}\varphi\,\partial_{\nu}\varphi  + \frac{3}{4}\frac{1}{\varphi^{2}} g_{\mu\nu}\,\partial^{ \lambda } \varphi\partial_{ \lambda}\varphi -\frac{1}{4}g_{\mu\nu}\left(\frac{\Sigma}{\varphi} + F(\Sigma)\right)    \\
  & &  +   \frac{1}{\varphi}\widetilde{\nabla}_{\mu}\partial_{\nu}\varphi  -
  \frac{1}{\varphi}g_{\mu\nu}\,\widetilde{\square}\varphi\,.   \label{4.12}
\end{eqnarray}

The RHS of this equation can be  rearranged as the sum of  the
energy-momentum tensors of standard perfect fluid matter and  of a
scalar field being
\begin{eqnarray}
\nonumber G_{\mu\nu}(\{\}) &=& \frac{1}{\varphi} \Sigma_{\mu\nu}-\frac{3}{2\varphi^{2}}\left[ \partial_{\mu}\varphi\,\partial_{\nu}\varphi - \frac{1}{2} g_{\mu\nu}\,\partial^{ \lambda } \varphi\partial_{ \lambda}\varphi +g_{\mu\nu}V(\varphi) \right. \\
  & & -\left.  \frac{2}{3}\varphi\widetilde{\nabla}_{\mu}\partial_{\nu}\varphi   +
  \frac{2}{3}g_{\mu\nu}\varphi\,\widetilde{\square}\varphi \right]\,, \label{4.12C}
\end{eqnarray}

where  the scalar field effective potential is  given by
\begin{equation}\label{4.13}
V(\varphi)= \frac{1}{6}\left(\varphi^{2}(f')^{-1}(\varphi)+\varphi
F^{-1}((f')^{-1}(\varphi))\right)
\end{equation}
and the operator  $\ \widetilde{\square}\equiv
\widetilde{\nabla}_{\lambda} \partial^{\lambda}$ is defined, as
above, by the covariant derivative without contorsion. This result
shows that our approach is fully equivalent  to  an effective
theory with perfect fluid matter and a scalar field with a
geometrical interpretation given by  Eq.(\ref{4.11}) and dynamics
given by the self-interaction potential  (\ref{4.13}). We obtained
this picture {\it without} adopting conformal transformation but
only splitting torsion from Christoffel contributions in the
connection. In other words, the result strictly depends on the
intrinsic non-linearities of the gravitation interacting with
perfect-fluid matter assuming a  gravitational Lagrangian density
of the form $\sqrt{-g}f(R)$ defined on a $U_{4}$ manifold.

\section{Field equations in presence of matter and spin}
As it is well known, spin can be one of the sources of torsion. In
1923,  Cartan showed that intrinsic angular momentum (after called
the spin) could have an important role in a geometric theory of
space-time like General Relativity \cite{cartan}. He showed that
spin could originate torsion. This idea was considered also by
Sciama \cite{sciama} and Kibble \cite{kibble} and, in more
recently, by Hehl and his coworkers \cite{hehl} and Trautmann and
his coworkers \cite{trautmann}. The path followed in this stream
of research was to extend GR to a theory
(Einstein-Cartan-Sciama-Kibble, ECKS, theory) in which spin is the
source of torsion. In the limit of zero spin distribution, the
theory reduces to standard GR.

In the ECSK theory,  spin and torsion are related by an algebraic
equation  and  torsion does not propagate. In this case, only
considering  a distribution of aligned spins, described by a
spin-density tensor $\tau_{\sigma\mu}^{\phantom{\sigma\mu}\rho}$,
it is possible to deal with  torsion as a fluid. It is then
necessary not only to have fluid of particles with spin, but also
the average spin alignment has to be different from zero, in order
to produce torsion. To deal with spin-torsion cosmology in the
ECKS approach, it is necessary to consider a cosmological fluid
where spin alignment is produced, for example, by huge magnetic
fields. Such field are difficult to be found  in standard
cosmological situations as discussed in several works in
literature.

In \cite{Capozziello:2001mq}, it was discussed in details  that
spin is not the only source of torsion and that the torsion tensor
can be decomposed in three irreducible tensors: $i)$ the totally
antisymmetric torsion, $ii)$ the vectorial torsion and $iii)$ the
traceless torsion.   It was shown, for example, that a spin field
generates a totally antisymmetric torsion, while a classical spin
fluid, as that introduced by Mathisson \cite{mathisson} and
Weyssenhoff \cite{weyssenhoff}, generates a traceless torsion.

In the previous section of this paper, we have seen  that the
vectorial part of torsion can be generated by the logarithmic
derivative of a given scalar field $\varphi$ of geometric origin.

Considering also spin in the context of this paper  leads us to
have a combined effect of different sources of torsion. In other
words, one has to take into account the vectorial torsion,
obtained by the covariant derivative of $\varphi$, and  the
classical spin average effect.  If in the action (\ref{2.1}), we
take into account a material Lagrangian $\mathcal{L}_{m}$ with
spin, the field equations have to be modified in the following way

\begin{equation}\label{msp1}
    \frac{\delta \sqrt{-g}f(R)}{\delta
    g_{\mu\nu}}=- f'(R) R^{\mu\nu} + \frac{1}{2} g^{\mu\nu}f(R)+
    \breve{\nabla}_{\lambda}
    (U^{\mu\nu\lambda}-U^{\nu\lambda\mu}+U^{\lambda\mu\nu})=
    \frac{1}{2}\frac{1}{\sqrt{g}}\frac{\delta \mathcal{L}_{m}}{\delta
    g_{\mu\nu}}\equiv\Sigma^{\mu\nu}\,,
\end{equation}

\begin{equation}\label{msp2}
   U_{\sigma\mu}^{\phantom{\sigma\mu}\rho}\equiv  \frac{\delta \sqrt{-g}f(R)}{\delta
    K_{\alpha\beta}^{\phantom{\alpha\beta}\delta}}=g^{\mu\nu}\partial_{\lambda} (f'(R))
 \left(\delta^{\lambda}_{\mu}\delta^{\rho}_{\sigma}-\delta^{\lambda}_{\sigma}\delta^{\rho}_{\mu}\right)+
2f'(R)g^{\mu\nu}T_{\sigma\mu}^{\phantom{\sigma\mu}\rho} =
\frac{1}{\sqrt{g}}\frac{\delta \mathcal{L}_{m}}{\delta
K_{\rho}^{\phantom{\varrho}\sigma\mu}}
\equiv\tau_{\sigma\mu}^{\phantom{\sigma\mu}\rho}\,.
\end{equation}
Such equations result more complicated than Eqs.(\ref{2.2}),
(\ref{2.3}). In this paper, we will not consider anymore the
presence of the spin fluid as one of the source of torsion. The
remarks of this section have been necessary in order to point out
that torsion can have also a purely geometric origin. In this
case, despite of the spin fluid which gives only an algebraic
relation, the torsion dynamical field related to the geometric
part can greatly contribute to the cosmic dynamics being a very
natural source for cosmic speed up.

\section{ The weak and strong energy conditions in $f(R)$ gravity with torsion}

As we said, the $f(R)$ gravitational field equations with torsion
can be expressed in the equivalent  Einstein-like form

\begin{equation}\label{5.1}
    G_{\mu\nu}= \frac{1}{\varphi} \Sigma_{\mu\nu} - \frac{3}{2}\frac{1}{\varphi^{2}} \sigma_{\mu\nu}
\end{equation}
where $\sigma_{\mu\nu}$ is the effective  scalar field
energy-momentum tensor in Eq.(\ref{4.12C}). Such a term has a
purely geometric origin and takes in account  the nonminimal
coupling between standard fluid matter and $f(R)$ gravity. It is
important to stress that it gives a  negative contribution  in RHS
of Eq. (\ref{5.1}) and it is possible to show that it can
naturally give rise to the observed accelerated behavior (due to
$\Omega_{\Lambda}\sim 0.7$) and the clustering properties (due to
$\Omega_{m}\sim 0.3$) of the standard $\Lambda$CDM.

In order to consider the properties of  $\sigma_{\mu\nu} $  and
its relevance for the cosmic evolution, let us recast it as a
(perfect) fluid energy-momentum tensor. Following this
prescription,  after considering Eq.(\ref{4.10}), let us  define

\begin{equation}\label{5.2}
     U^{\mu}= \frac{T^{\mu}}{\sqrt{T^{\lambda}T_{\lambda}}}
\end{equation}

which plays the role of a {\it four-velocity}. We are assuming
that $T^{\lambda} $ is a timelike-vector. Then the {\it energy
density} is

$$\rho=\frac{1}{\varphi }\varrho  - \frac{3}{2}\frac{1}{\varphi^{2}}\sigma_{\mu\nu}U^{\mu}U^{\nu}=\frac{1}{\varphi }\varrho- \frac{3}{2}\frac{1}{\varphi^{2}} \left( \frac{1}{2}  \partial_{\lambda}\varphi \partial^{\lambda}\varphi+ V(\varphi) +  \frac{2}{3} \varphi\,\widetilde{\square}\varphi \right.$$
\begin{equation}\label{5.3}
  \left. -\frac{2}{3}\varphi( \partial_{\lambda}\varphi \partial^{\lambda}\varphi)^{-1}
  \partial^{\mu}\varphi\partial^{\nu}\varphi\widetilde{\nabla}_{\mu}\partial_{\nu}\varphi\right)\,,
\end{equation}
and the {\it pressure}
\begin{equation}\label{5.4}
     P= \frac{1}{\varphi }p - \frac{3}{2}\frac{1}{\varphi^{2}} \left[\frac{1}{2}
     \partial_{\lambda}\varphi \partial^{\lambda}\varphi- V(\varphi) -
     \frac{2}{3}\left(\frac{2}{3} \varphi\,\widetilde{\square}\varphi+\frac{1}{3}\varphi
     ( \partial_{\lambda}\varphi \partial^{\lambda}\varphi)^{-1}\partial^{\mu}\varphi
     \partial^{\nu}\varphi\widetilde{\nabla}_{\mu}\partial_{\nu}\varphi
     \right)\right]\,.
\end{equation}
For our purposes, it is important  to consider the weak and strong
energy conditions \cite{ellis}. The weak energy condition implies
that
\begin{equation}\label{5.5}
     \rho \geq 0\ \ \ \Rightarrow \varrho\geq  \frac{3}{2}\frac{1}{\varphi}\left[\frac{1}{2}  \partial_{\lambda}\varphi \partial^{\lambda}\varphi+ V(\varphi) +  \frac{2}{3} \varphi\,\widetilde{\square}\varphi
 -\frac{2}{3}\varphi( \partial_{\lambda}\varphi \partial^{\lambda}\varphi)^{-1}\partial^{\mu}\varphi\partial^{\nu}\varphi\widetilde{\nabla}_{\mu}\partial_{\nu}\varphi\right]
\end{equation}
where $\varrho=\Sigma_{\mu\nu}U^{\mu}U^{\nu}$ and
$p=\Sigma_{\mu\nu}(g^{\mu\nu}+U^{\mu}U^{\nu})$ are respectively
the energy density and the pressure for the standard-fluid matter
source.

The strong energy condition holds, if the RHS tensor of Eq.
(\ref{5.1}) satisfies the condition
 \begin{equation}\label{5.6}
      \left[\frac{1}{\varphi} \Sigma_{\mu\nu}  - \frac{3}{2}\frac{1}{\varphi^{2}} \sigma_{\mu\nu}  -\frac{1}{2}\left(\frac{1}{\varphi} \Sigma - \frac{3}{2}\frac{1}{\varphi^{2}} \sigma \right)  g_{\mu\nu}\right]U^{\mu}U^{\nu}=\frac{1}{2}\left(\rho +3P \right)\geq0
\end{equation}
corresponding to
\begin{equation}\label{5.7}
 \varrho+3p\geq  \frac{3}{2}\frac{1}{\varphi} \left[
   \partial_{\lambda}\varphi \partial^{\lambda}\varphi- V(\varphi) -\frac{2}{3}\left(\frac{1}{2} \varphi\,\widetilde{\square}\varphi+ \varphi( \partial_{\lambda}\varphi \partial^{\lambda}\varphi)^{-1}\partial^{\mu}\varphi\partial^{\nu}\varphi\widetilde{\nabla}_{\mu}\partial_{\nu}\varphi \right)\right].
\end{equation}
When this condition is  satisfied, the universe undergoes  a
decelerated cosmological expansion, viceversa  an accelerated
expansion is expected when it is violated.  These inequalities
strictly depend on the form of the function $f(R)$ and on the
nonminimal coupling of $f(R)$ with the standard matter given by
$\Sigma$ (see Eqs.(\ref{2.4}) and (\ref{4.11})).

In conclusion, we can say that introducing  torsion of geometric
origin  in $f(R)$ gravity can lead to an accelerated behavior of
the  universe due to  a repulsive nonlinear interaction of the
(baryon) matter with itself. A qualitative discussion of this
point is given in \cite{cianci1}. There is discussed also the
possibility that comparing the effective ${\displaystyle
\Sigma_{eff}=\frac{\Sigma}{\varphi}}$ and ${\displaystyle
V_{eff}=\frac{V(\varphi)}{\varphi^2}}$ can lead to comparable
values for $\Omega_m$ and $\Omega_{\Lambda}$. This could be a
straightforward way to solve the so-called {\it coincidence
problem} \cite{LambdaTest,LambdaRev,PR03,copeland}.

 \section{The cosmological solutions in presence of perfect fluid matter}

In order to determine  the cosmological equations in presence of
perfect-fluid matter, we can arrange   Eq.(\ref{4.12})in the form
$$R_{\mu}^{\nu}=\frac{1}{\varphi}\Sigma_{\mu}^{\nu}
-\frac{1}{4}\left(\frac{1}{\varphi}\Sigma-F(\Sigma)\right)\delta_{\mu}^{\nu}-\frac{3}{2}\frac{\partial_{\mu}\varphi \partial^{\nu}\varphi }{\varphi^{2}}  $$
\begin{equation}\label{4.a1}
    + \frac{\partial^{\nu}\partial_{\mu}\varphi}{\varphi}+
\frac{1}{2}\frac{\partial_{\lambda}\partial^{\lambda}\varphi}{\varphi}
\delta_{\mu}^{\nu}- g^{\nu\lambda}\Gamma _{\mu\lambda}^{\sigma
}\frac{\partial_{\sigma}\varphi}{\varphi} + \frac{1}{2}
\Gamma_{\lambda\sigma}^{\lambda}\frac{\partial^{\sigma}\varphi
}{\varphi} \delta_{\mu}^{\nu}\,.
\end{equation}

 We  consider the components of the Ricci tensor

\begin{equation}\label{4.a2}
     R_{0}^{0}=\frac{1}{\varphi}\Sigma_{0}^{0}
- \frac{1}{4\varphi}\Sigma+\frac{1}{4}F(\Sigma) -\frac{3}{2}\frac{
\dot{\varphi}^{2} }{\varphi^{2}}+\frac{3}{2}\frac{
\ddot{\varphi}}{\varphi }+\frac{3}{2}H\frac{
\dot{\varphi}}{\varphi}
\end{equation}
and
\begin{equation}\label{4.a3}
  R_{1}^{1}=\frac{1}{\varphi}\Sigma_{1}^{1}
- \frac{1}{4\varphi}\Sigma+\frac{1}{4}F(\Sigma)
+\frac{\ddot{\varphi}}{2\varphi}+\frac{5}{2}H\frac{\dot{\varphi}}{\varphi}\,.
\end{equation}
In the homogeneous and isotropic FRW universe, we have
\begin{eqnarray}
R_{0}^{0}&=&-3\frac{\ddot{a}}{a}\,,\\
R_{1}^{1}&=&-\left(\frac{\ddot{a}}{a}+2\frac{\dot{a}^{2}}{a^{2}}+\frac{2k}{a^{2}}\right)\,,\\
\Sigma_{0}^{0}&=&\varrho\,,\\
\Sigma_{1}^{1}&=&-p=-(\gamma-1)\varrho\\
\Sigma&=&\varrho-3p=(4-3\gamma)\varrho\,,
\end{eqnarray}
where $0\leq \gamma\leq 1$ is the adiabatic index of the equation
of state. The cosmological equations are then
\begin{equation}\label{4.a5}
     -3\frac{\ddot{a}}{a}=\frac{3}{4}\frac{\gamma\varrho}{\varphi}+\frac{1}{4}F(\Sigma)-
     \frac{3}{2}\frac{ \dot{\varphi}^{2} }{\varphi^{2}}+\frac{3}{2}\frac{ \ddot{\varphi}}{\varphi }+
     \frac{3}{2}H\frac{ \dot{\varphi}}{\varphi}\,,
\end{equation}
\begin{equation}\label{4.a6}
  -\left( \frac{\ddot{a}}{a}+2\frac{\dot{a}^{2}}{a^{2}}+2\frac{k}{a^{2}}\right)=-\frac{\gamma}{4}\frac{\varrho}{\varphi}
 +\frac{1}{4}F(\Sigma)
 +\frac{1}{2}\frac{\ddot{\varphi}}{\varphi}+\frac{5}{2}H\frac{\dot{\varphi}}{\varphi}\,.
\end{equation}
By composing these two  equations to eliminate the $\ddot{a}/{a}$,
one obtains
\begin{equation}\label{4.a7}
    H^{2}+\frac{k}{a^{2}}=\frac{\gamma}{4}\frac{\varrho}{\varphi}-\frac{1}{12}F(\Sigma)-
\frac{1}{4}\frac{\dot{\varphi}^{2}}{\varphi^{2}}-H\frac{\dot{\varphi}}{\varphi}
\end{equation}
which, solved with respect to $H$, gives
\begin{equation}\label{4.a8}
    H=-\frac{\dot{\varphi}}{2\varphi}+\frac{\varepsilon}{2}\sqrt{\Delta_{k}}\,,
\end{equation}
with $\varepsilon=\pm 1$ and
\begin{equation}\label{4.a9}
    \Delta_{k}=\frac{\gamma\varrho}{\varphi}-\frac{1}{3}F(\Sigma)\,.
-\frac{4k}{a^{2}}
\end{equation}
In order to derive the evolution of the density $\varrho$, let us
consider the following equation, derived by combining
Eqs.(\ref{4.a5}) and (\ref{4.a6}),
\begin{eqnarray}
 \nonumber  \dot{H}+H^{2} &\equiv& -\frac{\ddot{\varphi}}{2\varphi}+\frac{\dot{\varphi}^{2}}{2\varphi^{2}}
+\frac{\varepsilon}{4}\frac{\dot{\Delta}_{k}}{\sqrt{\Delta_{k}}} + \frac{\dot{\varphi}^{2}}{4\varphi^{2}}-\frac{\varepsilon}{2}\sqrt{\Delta_{k}}\frac{\dot{\varphi} }{\varphi}+ \frac{1}{4}\Delta_{k}\\
    &=& - \frac{\gamma\varrho}{4\varphi}-\frac{1}{12}F(\Sigma)+\frac{ \dot{\varphi}^{2} }{2\varphi^{2}}-
    \frac{ \ddot{\varphi}}{2\varphi }-\frac{1}{2}H\frac{\dot{\varphi}}{\varphi}
\end{eqnarray}
This equation simplifies to
\begin{equation}\label{4.a14}
  \frac{\varepsilon}{4}\frac{\dot{\Delta}_{k}}{\sqrt{\Delta_{k}}}+
  \frac{\dot{\varphi}^{2}}{4\varphi^{2}}-\frac{\varepsilon}{2}\sqrt{\Delta_{k}}\frac{\dot{\varphi} }
  {\varphi}+ \frac{1}{4}\Delta_{k}=-\frac{\gamma\varrho}{4\varphi}-\frac{1}{12}F(\Sigma)-
  \frac{1}{2}H\frac{\dot{\varphi}}{\varphi}
\end{equation}
A further simplification comes from writing, explicitly in
Eq.(\ref{4.a14}), the expressions for $\Delta_{k}$ and $H$. We
finally find the evolution equation
\begin{equation}\label{4.a15}
     \frac{\varepsilon}{4}\frac{\dot{\Delta}_{k}}{\sqrt{\Delta_{k}}}=-\frac{\gamma\varrho}{2\varphi}+\frac{k}{a^{2}}+
     \frac{\varepsilon}{4}\sqrt{\Delta_{k}}\frac{\dot{\varphi}}{\varphi}
\end{equation}

These equations can  be solved by quadrature. The scale factor
$a(t)$ is obtained by solving  Eq.(\ref{4.a8}).

As an example, let us assume  $f(R)\propto R^{n}$ and consider a
perfect fluid with equation of state $p=(\gamma-1)\varrho$ in a
universe with $k=0$. After some calculations we find, from Eqs.
(\ref{4.5}) and (\ref{4.11}),
\begin{equation}\label{6.15b}
    F(\varrho)=\left(\frac{\Sigma}{n-2}\right)^{1/n}=\left(\frac{(4-3\gamma)\varrho}{n-2}\right)^{1/n}\,,
\end{equation}
\begin{equation}\label{6.15}
     \varphi(\varrho)=n\left(\frac{(4-3\gamma)\varrho}{n-2}\right)^{(n-1)/n}\,,
\end{equation}
and

\begin{equation}\label{6.16}
    \Delta_{0}(\varrho)=C_{1}\varrho^{1/n}\,,
\end{equation}
with
\begin{equation}\label{6.16C}
 C_{1}=-
 \frac{1}{3}\frac{-3\gamma+(4-3\gamma)^{1/n}(n-2)^{-1/n}C_{2}}{3C_{2}}\,.
\end{equation}

Solving  Eq.(\ref{4.a15}), we find
\begin{equation}\label{4.a17}
    \varrho(t)= \left( \frac{\gamma}{(n-2) C_{2}^{1/2}C_{1}} \right)^{-2n}(t-t_{0})^{-2n}
\end{equation}
where ${\displaystyle
C_{2}=n\left[\frac{(4-3\gamma)}{n-2}\right]^{(n-1)/n}}$. The
explicit  expression for $\varphi(t)$ is immediately found and
\begin{equation}\label{4.a18}
    \Delta_{0}(t)=
    \frac{(C_{1}^{2}C_{2}^{2}(n-2)^{2})}{\gamma^{2}}\frac{1}{(t-t_{0})^2}\,.
\end{equation}
The evolution  of $a(t)$ is obtained  by integrating
Eq.(\ref{4.a8}). We have solutions for $n\geq 4$, that is
\begin{equation}\label{4.a19}
    \frac{\dot{a}}{a}=\frac{n-1}{(t-t_{0})}+\varepsilon\frac{(C_{1}C_{2}(n-2))}{2\gamma}\frac{1}{(t-t_{0})}\,,
\end{equation}
and then
\begin{equation}\label{4.a20}
    a(t)=A_1 (t-t_{0})^{\alpha}\;\;\;\;\mbox{where}\;\;\;\;\alpha=n-1+\varepsilon\frac{(C_{1}C_{2}(n-2))}{2\gamma}\,.
   \end{equation}
This is a power law expansion  depending on the parameters
$\gamma$ and $n$. Accelerated/decelerated  behaviors are easily
achieved by discussing the values of these parameters with respect
to the Bianchi identities. As an example, let us consider the case $n=4$. The decelerated behavior is obtained for
\begin{equation}
0<\alpha=3+\varepsilon\frac{C_1C_2}{\gamma}<1\,,
\end{equation}
otherwise the universe is accelerating. The transition from the decelerated  to the accelerated phases is achieved for $\alpha=1$  and then the redshift has to evolve as $z\sim t^{-1}$.
Following \cite{Hu}, it is easy to see that the condition $[1+\Omega_{eff}]\simeq 0.2$ to recover $\Lambda$CDM prescriptions is easily recovered depending on the couple of parameters $\{n,\gamma\}$.

\section{Discussion and Conclusions}
Several issues from cosmology and astrophysics are telling us that
the Einstein General Relativity should be revised in order to
avoid  shortcomings as dark energy and dark matter which, up to
now, have not been probed at any fundamental level, but manifest
their presence at large scales. Despite of the lack of final
experimental evidences, the {\it dark side} of the universe should
constitute almost the 95 \% of the whole matter-energy content.

This situation is extremely disturbing and a more "economic" way
to solve the problem could be to revise the geometric part of the
gravitational interaction  extending the GR. In $f(R)$ gravity,
this point of view is considered and, in general,  the good
results of the Einstein theory are  preserved  trying, at the same
time,  to encompass the  observational data in some
self-consistent scheme \cite{GRGrev}.

In this paper, we have pushed forward this approach taking also
into account  the role of torsion in $f(R)$ cosmology. With
respect to the ECKS theory, where torsion couples algebraically to
the spin and matter without dynamics, we have put in evidence the
fact that torsion fields of purely geometric origin can perfectly
mimic the role of scalar fields in cosmological evolution. Such a
feature naturally gives rise to accelerated behaviors as in dark
energy models.   In general, torsion has no effect in vacuum but,
for $R^2$, it is a source for the field equations leading to de
Sitter-like expansions.

In presence of standard fluid matter, torsion and curvature
degrees of freedom give rise to nonminimal couplings which
determine cosmological evolution. This  depends on the form of
$f(R)$. Also in these cases, accelerated behaviors are easily
achieved.
However, also in presence of torsion,
$f(R)$-gravity models must satisfy some viability conditions in
order to describe cosmic acceleration. Such conditions  are summarized
\cite{pogosian}. The cosmological models proposed here fully satisfy these
compatibility conditions since the presence of torsion acts as a
further additive fluid with the only effect to move the extremum of the
effective potential which, following the notation in \cite{pogosian},
lies at the GR value $R\sim(\rho-3p)$.

As a final comment, it is worth noticing that we have not used any
conformal transformation so the issue to choose between the
Einstein and the Jordan frame is avoided. Besides, torsion
naturally results as a fundamental scalar field whose origin is
perfectly understood.

In a forthcoming paper, we will discuss more realistic $f(R)$
gravity models with torsion confronting them with data.


\begin{thebibliography}{99}


\bibitem{LambdaTest} U. Seljak  et al.  Phys. Rev. \textbf{D71},
103515 (2005).

\bibitem{LambdaRev} S.M. Carroll, W.H. Press, E.L. Turner,   Ann. Rev.
Astron. Astroph. \textbf{30}, 499 (1992).

\bibitem{PR03} Peebles P.J.E., Rathra B., Rev. Mod. Phys. \textbf{75}, 559 (2003);
Padmanabhan T., Phys. Rept., \textbf{380}, 235 (2003).

\bibitem{copeland} E.J. Copeland, M. Sami, S. Tsujikawa, Int. J. Mod.
Phys. \textbf{D 15}, 1753 (2006).

\bibitem{will} C.~M.~Will,
%``The confrontation between general relativity and experiment,''
Living Rev. Relativity \textbf{9} (2006), arXiv:gr-qc/0510072.

\bibitem{kleinert} H. Kleinert, H.-J. Schmidt, Gen. Relativ.
Grav. \textbf{34} 1295 (2002).

\bibitem{noi} S. Capozziello, Int. J. Mod. Phys. \textbf{D 11}, 483 (2002).\newline
S. Capozziello, S. Carloni, A. Troisi, Rec. Res. Dev. in Astron.
and Astroph. \textbf{1}, 1 (2003)
(arXiv\,:\,astro\,-\,ph/0303041).\newline S.D. Odintsov, S. Nojiri
  Phys. Lett. \textbf{B 576}, 5 (2003).\newline
  S. Capozziello, V.F. Cardone, S. Carloni, A. Troisi,  Int. J. Mod. Phys. \textbf{D 12},
1969 (2003).\newline S.M. Carroll, V. Duvvuri, M. Trodden, M.
Turner, Phys. Rev. \textbf{D 70}, 043528 (2004).
\newline G. Allemandi, A. Borowiec, M. Francaviglia,  Phys. Rev. \textbf{D 70},
103503 (2004).
\newline
S. Nojiri and S.D. Odintsov, Phys. Rev. \textbf{D } 0307288
\newline
S. Nojiri  and S.D. Odintsov,  Gen. Rel.
Grav. \textbf{36}, 1765 (2004).
\bibitem {odirev} S.Nojiri and S.D. Odintsov, Int. J. Meth. Mod. Phys.
\textbf{4}, 115 (2007).

\bibitem{GRGrev} S. Capozziello and M. Francaviglia, Gen. Rel. Grav. {\bf 40}, 357
(2008).

\bibitem{faraoni} T. P. Sotiriou, V. Faraoni,
arXiv:0805.1726, (2008).


\bibitem{noipla} S. Capozziello, V.F. Cardone, S. Carloni, A. Troisi,
Phys. Lett. \textbf{A 326}, 292 (2004).

\bibitem{mond} M. Milgrom,   Astroph. Journ.,\textbf{270}, 365 (1983); \newline J. Bekenstein,
Phys. Rev. \textbf{D 70}, 083509 (2004).

\bibitem{jcap}S. Capozziello, V.F. Cardone and A. Troisi JCAP \textbf{08},
001 (2006).

\bibitem{mnras}S. Capozziello, V.F. Cardone, A. Troisi,
Mon.\ Not.\ Roy.\ Astron.\ Soc.\ \textbf{375}, 1423 (2007).

\bibitem{sobouti}Y. Sobouti, A\&A, \textbf{464}, 921 (2007).

\bibitem{salucci} C.Frigerio Martins
and P. Salucci, MNRAS \textbf{381}, 1103, (2007).

\bibitem{mendoza}S. Mendoza and Y.M. Rosas-Guevara, A\&A, \textbf{472}, 367 (2007).

\bibitem{instabilities-f(R)}V.~Faraoni, Phys.\ Rev.\ \textbf{D 72}, 124005
(2005); \newline G.~Cognola and S.~Zerbini, J.\ Phys.\ \textbf{A
39}, 6245 (2006); \newline G. Cognola, M. Gastaldi and S. Zerbini,
arXiv: gr\,-\,qc/0701138.

\bibitem{ghost-f(R)}K.~S.~Stelle, Gen.\ Rel.\ Grav.\ \textbf{9}, 353 (1978).

\bibitem{mimicking} S. Capozziello, V.F. Cardone, A. Troisi,  Phys. Rev.
\textbf{D 71}, 043503 (2005).

\bibitem{Hu} W.~Hu and I.~Sawicki,
%``Models of f(R) Cosmic Acceleration that Evade Solar-System Tests,''
Phys.\ Rev.\ D \textbf{76}, 064004 (2007).

\bibitem{Star}A.~A.~Starobinsky,
%``Disappearing cosmological constant in f(R) gravity,''
JETP Lett.\ \textbf{86}, 157 (2007).

\bibitem{Odintsov1}
S.~Nojiri and S.~D.~Odintsov, Phys.\ Lett.\  B {\bf 652}, 343
(2007).

\bibitem{tsuji} S. Capozziello, S. Tsujikawa, Phys. Rev.  {\bf D 77}, 107501 (2008).

\bibitem{prado}
S. Capozziello, P. Martin-Moruno, C. Rubano,  Phys. Lett.
\textbf{B 664}, 12 (2008).

\bibitem{antonio}
S. Capozziello and A. De Felice,  Jou. Cosm. Astrop. Phys.  {\bf 08}, 016,  2008
B. Vakili, arXiv: 0804.3449 [gr-qc] (2008).

\bibitem{hehl}
  F.~W.~Hehl, P.~Von Der Heyde, G.~D.~Kerlick and J.~M.~Nester,
  %``General Relativity With Spin And Torsion: Foundations And Prospects,''
  Rev.\ Mod.\ Phys.\  {\bf 48}, 393 (1976).



\bibitem{Capozziello:2001mq}
  S.~Capozziello, G.~Lambiase and C.~Stornaiolo,
Annals.  Phys. (Leipzig)   {\bf 10}, 713 (2001).


\bibitem{cianci1}
S. Capozziello, R. Cianci, C. Stornaiolo, and S. Vignolo, Class.
Quant. Grav. {\bf 24}, 6417 (2008).

\bibitem{cianci2}
S. Capozziello, R. Cianci, C. Stornaiolo, and S. Vignolo,
 Int. Jou. Geom. Meth. Mod. Phys. \textbf{5}, 765 (2008).

\bibitem{magnano}
G. Magnano, M. Ferraris, and M. Francaviglia,
Gen.Rel.Grav. \textbf{19}, 465 (1987).


\bibitem{gasperini}
M. Gasperini, Gen. Rel. Grav. \textbf{30}, 1703 (1998).

\bibitem{szydlowski}
M. Szydlowski and A. Krawiec, Phys. Rev. \textbf{D 70}, 043510
(2004).

\bibitem{cartan}
E. Cartan, Ann. Sci. Ec. Normale Super. \textbf{40}, 325 (1923).


\bibitem{sciama} D.W. Sciama,  {\it On the analog between charge and spin in
General Relativity}, in Recent Developments in General Relativity,
Festschrift for Leopold Infeld, (1962) 415, Pergamon Press, New
York.

\bibitem{kibble} T.W. Kibble,
 J. Math. Phys. \textbf{2}, 212 (1960).


\bibitem{trautmann}
 A. Trautmann, Bull. Acad. Pol. Sci., Ser. Sci.,
Math., Astron. Phys. \textbf{20}, 895 (1972).


\bibitem{mathisson} M. Mathisson, Acta Phys. Pol. \textbf{6}, 163
(1937).

\bibitem{weyssenhoff} J. Weyssenhoff and A. Raabe, Acta Phys. Pol.
\textbf{9}, 7 (1947).

\bibitem{ellis}
S.W. Hawking and G.F.R. Ellis, {\it The Large Scale Structure of
Space-Time}, Cambridge Univ. Press, Cambridge (1973).

\bibitem{Pogosian} L.~Pogosian and A.~Silvestri
Phys.\ Rev.\ D \textbf{77}, 023503 (2008).


\end{thebibliography}
\end{document}